\magnification 1200
\nopagenumbers
\input boxedeps.tex
\input boxedeps.cfg
\SetDefaultEPSFScale{600}
\baselineskip=7mm
\hsize=5.0in

\centerline{\bf{Properties of Solutions in 2+1 Dimensions}}
\vskip .2truein
\centerline{\it{  Dedicated to Prof.Yavuz Nutku on his sixtieth birthday}}
\vskip .7truein
\centerline {M.Horta\c csu $^{* \dagger }$, Hasan Tuncay \" Oz\c celik $^{*}$, 
Bar\i \c s Yap\i \c skan $^{*}$ }
\vskip .1truein
\centerline { $^{*} $ Physics Department, Faculty of Science and Letters}
\vskip .03 truein
\centerline { ITU 80826, Maslak, Istanbul, Turkey}
\vskip .1truein
\centerline {$^{\dagger}$ Feza G\" ursey Institute, Istanbul, Turkey}
\vskip 1.3truein
\centerline {\bf {Abstract}}
\noindent
We solve the Einstein equations for the 2+1 dimensions with and without scalar fields. We calculate the entropy, Hawking temperature and the emission probabilities for these cases. We also compute the Newman-Penrose coefficients for different solutions and compare them.

\vskip .3truein
\noindent
key words: BTZ solutions, thermodynamical properties, Newman-Penrose coefficients
\vskip 1.5truein
\noindent
hortacsu@itu.edu.tr
\baselineskip=18pt
\footline={\centerline{\folio}}
\pageno=1
\vfill\eject

\noindent
{\bf{INTRODUCTION}}
\vskip.2truein
It is  common practice to study similar phenomena in lower dimensions when 
calculations in models in realistic physical dimensions are too hard to 
perform. For this reason toy models are used in quantum field theory and 
particle physics where physics in lower dimensions is studied. [1] 

This was not done in general relativity, though, for a long time. 
One reason for this may be because the study of general relativity in 
(2+1) dimensions was thought to be trivial.  The degree of freedom of a 
graviton is given by (d-3) in any number of dimensions. In (2+1) dimensions, 
$d=3$ and this count does not allow any local degrees of freedom to the related 
field. One could study models only with interesting global properties in this 
dimension. Furthermore, in  2+1 dimensions the Riemann tensor does not have any 
other components as those given by the Ricci tensor, i.e. the Weyl tensor which 
gives half of the components of the Riemann tensor in $d=3+1$ does not exist 
anymore.  This fact makes it impossible to have a Ricci flat space with non-
trivial Riemann tensor components.

This is why a "black-hole" solution of the vacuum theory, given by Banados, 
Teitelboim and Zanelli (BTZ), [2],  was a surprise to the general relativity 
community.
This solution approached not to the Minkowski, but to the anti-de Sitter space 
asymptotically. The Einstein equations were modified with respect to the (3+1) 
case by the introduction of the cosmological constant.  These facts did not 
make the BTZ solution less interesting.

Actually (2+1) dimensional gravity was studied by Deser, Jackiw, 't Hooft and 
Templeton [3] much before. They had added a Chern-Simons term to the Riemann-
Hilbert action and found solutions to the problem. Still the BTZ solution came 
as an unexpected event.

BTZ studied also the rotating and the charged "black holes" in their original 
paper. An error was pointed out to their treatment of the both rotating and 
charged case [4,5,6]. The authors revised their solution in a later paper [7].

Later studies were made of the case of coupling of a scalar field to the 
gravitational field [8]. Among additional references on this field we can cite the papers by Virbhadra, Dias and Lemos [9). Here we will derive these known solutions using 
slightly different methods. We will also try to treat the case when the scalar 
field is time dependent. This solution can not be written in an analytical 
form, since the resulting function can not be inverted. We will give only the 
numerical solution for this case.

We will, then, study the thermodynamical properties of the solutions using the 
methods of Kraus, Keski-Vakkuri and F.Wilczek [10,11,12]. In the final part we 
calculate the coefficients in the Newman-Penrose formalism [13] for the BTZ 
solution given in reference 7 and point to the similarities with the 
coefficients with the "wrong" solution proposed earlier.

The details of the calculations of  second and the third sections are given in the Thesis prepared by Hasan Tuncay \" Oz\c celik (ITU 2002). The fourth section is partially based on the Thesis prepared by Bar\i \c s Yap\i \c skan (ITU 2000).

\vskip.5truein
\noindent
{\bf{Solutions}}
\vskip.2truein
We start by writing the Einstein equations of motion
$$ G_{\mu \nu} + g_{\mu \nu} \Lambda = k T_{\mu \nu} \eqno{1} $$
where
$$ G_{\mu \nu} = R_{\mu \nu}- {{1}\over{2}} g_{\mu \nu}, \eqno{2} $$
$\Lambda$ is the cosmological constant, $ R_{\mu \nu}$ is the Ricci tensor, $k= 
{{8\pi G}\over{c^2}}$ where $G$ is the Newton's gravitational constant and $T_
{\mu \nu} $ is the stress-energy tensor. Here we use the convention that a 
positive value for $\Lambda$ will denote the anti-de Sitter space, and use the 
units where numerically $G=c=1$.
\vskip.1truein
\noindent
I. The non-rotating and uncharged black-hole

\noindent
We take the metric as 
$$ ds^2 = -v(r) dt^2 + w( r) dr^2 +r^2 d \theta ^2 . \eqno{3} $$
The Einstein equations read 
$$ T_{1}^{1}= \Lambda + {{w'}\over {2rw^2}}, \eqno{4} $$
$$ T_{2}^{2}= \Lambda - {{v'} \over {2rw^2}}, \eqno{5} $$
$$ T_{3}^{3}= {{1}\over {4v^2 w^2}} (4\Lambda w^2 +wv'^2 +vv'w' -2vwv'') . 
\eqno {6} $$
In these equations $'$ denotes differentiation with respect to r. In the vacuum 
solution all the components of the stress-energy tensor are zero. Using this 
fact, we can integrate equation 4 which  gives 
$$ w (r) =(\Lambda r^2 -C)^{-1} . \eqno {7} $$
This solution is used in equation 5 to give 
$$ v(r) =(\Lambda r^2 -C) . \eqno{8} $$ In these expressions $C$ is a constant 
which we will identify with the ADM mass $M$.
Equation 6 verifies that  the expressions found for $v ( r ) $ and $w( r ) 
$ are correct.

The resulting metric is

$$ ds^2 = -(\Lambda r^2 - M) dt^2 + (\Lambda r^2 -M)^{-1} dr^2 +r^2 d \theta 
^2. \eqno{9} $$

\vskip.1truein
\noindent
II. Rotating Black Hole

\noindent
We take the metric in the form

$$ ds^2 = -v(r) dt^2 + w( r ) dr^2 +J dt d\theta +r^2 d\theta^2 . \eqno{10} $$
Writing the Einstein equations and noting the fact that  all the components 
of the stress-energy tensor are zero, these equations can be solved 
straightforwardly.

The end result is that 
$$w( r ) = \left( \Lambda r^2 + {{J^2}\over {r^2}} -M \right)^{-1} , \eqno {11} 
$$
$$v( r ) = \Lambda r^2 - M. \eqno{12} $$
and the metric reads
$$ ds^2 = - \left( \Lambda r^2 + {{J^2}\over {r^2}} -M \right) dt^2 + \left( 
\Lambda r^2 + {{J^2}\over {r^2}} -M \right)^{-1} dr^2 + \left( rd\theta + {{J}
\over {r}} dt \right)^2 . \eqno{13} $$
Here the constant we found upon integrating the differential equations is 
interpreted as negative of the ADM mass. Upon equating the expression $\left( 
\Lambda r^2+{{J^2}\over{r^2}}-M \right) $ to zero, we can find the outer 
and the inner horizons $r_{+}$ and $r_{-}$ .
\vskip.1truein
\noindent
III. Charged Black Hole

\noindent
Here we equate the diagonal components of the stress-energy tensor to
$$ T_1^1 = T_2^2 = -T_3^3 = {{Q^2}\over{r^2}}. \eqno{14} $$ 
and choose the metric to be diagonal . Straightforward integration of the 
equations give the expression for the metric as 
$$ ds^2 = -(\Lambda r^2 - M -2Q^2 \log {{r}\over{r_{+}}}) dt^2 + (\Lambda r^2 - 
M -2Q^2 \log {{r}\over{r_ {+}}})^{-1} dr^2 +r^2 d\theta^2 . \eqno {15} $$
Here $r_{+} = \sqrt{{{M} \over { \Lambda}}} $.

As shown in references 4, 5 and 6, one can not obtain a charged rotating 
solution through these straightforward methods. Although the naive solution of 
the Einstein equations in this case give 

$$ ds^2 =-\left( v( r ) + {{J ( r )^2}\over {r^2}} \right) dt^2 + {{1}\over{v( 
r )}}dr^2 + \left( rd\theta +{{J( r ) }\over{r}}dt \right)^2 , \eqno {16} $$
where 
$$ v( r ) = \Lambda r^2 + {{J^2}\over{r^2}} -M -2B^2 \log {{{r}\over{r_{+}}}}, 
\eqno {17}$$ 
$$ r_{+}= \left( {{M+\sqrt{M^2-4J^2\Lambda}}\over{2\Lambda}}\right)^{{{1}\over
{2}}} \eqno {18} $$
the Maxwell equations, $\nabla^{\mu} F_{\mu \nu} = 0$, require 
$ JB=0 $.
If we take one of these constants equals to zero, we get back our previous 
solutions for the rotating uncharged or the charged non-rotating cases.
\vskip.1truein
\vfill\eject
\noindent
IV. Coupling to a Static Scalar Field

\noindent
Now we choose the stress-energy tensor in the conformal fashion.
$$ T_{\mu \nu } = (1-2\xi) \partial_{\mu} \phi( r ) \partial_{\nu}\phi( r ) + (2
\xi-1/2) g_{\mu \nu} g^{\lambda \tau} \partial_{\lambda} \phi( r ) \partial _
{\tau} \phi( r ) $$
$$- 2 \xi \phi( r )\left(\partial_{\mu \nu} \phi( r ) - \Gamma ^{\lambda}_{\mu 
\nu} \partial_{\lambda} \phi( r ) \right) +{{2}\over{d}} \xi g_{\mu \nu} \phi( 
r ) g^{\mu \nu} \phi( r );_{\mu \nu} $$
$$ - \xi \phi( r )^2 \left( R_{\mu \nu} - R g_{\mu \nu} \left( 1-4\xi {{d-1}
\over {d}} \right) \right). \eqno {19} $$
$d$ is the space-time dimension which is three in our case. 
$$ \xi= {{1}\over {4}} {{d-2}\over{d-1}}. \eqno {20} $$
The action of the d'Alembertian on a scalar field reads
$$ g^{\mu \nu} \phi_{; \mu \nu} = {{1}\over {\sqrt{-g}}} \left( \partial_
{\lambda} \sqrt{-|g|} g^{\lambda \tau} \partial_{\tau} \phi( r ) \right) . 
\eqno {21} $$
which we use in the equation of motion of the scalar field
$$ g^{\mu \nu} \phi_{; \mu \nu} +\xi \phi R =0 . \eqno {22} $$
We take the metric to be diagonal.

We find 
$$ ds^2= -v(r) dt^2+ {{1}\over{v( r )}} dr^2 +r^2d\theta^2 \eqno{23} $$ 
where 
$$ v( r ) =\left( {{(r-2B)(B+r)^2\Lambda}\over{r}}\right) \eqno{24} $$ and 
$$ \phi( r ) = \sqrt{ {{B}\over { \pi(r+B)}}}. \eqno {25} $$
$B$ is a constant.
This solution, also found in reference 8, gives 
$$ T_1^1 = T_2^2 = -{{1}\over{2 }} T_3^3 = -{{B^3\Lambda}\over {r^3}}. \eqno 
{26} $$ 

If we look for a time dependent solution for the scalar field, the metric 
becomes
$$ ds^2= -\left( 4r^2 \Lambda - {{2a(t)^3 C}\over{3r}}+ a(t) \right) +2 dr dt + 
r^2 d\theta^2 , \eqno {27} $$
and the scalar field is given as 
$$ \phi(r,t) = \sqrt {{{ a(t)^2C}\over{\pi(r+a(t)^2 c)}}} . \eqno {28} $$
Here $C$ is a constant and $a(t)$ is the solution of the differential equation
$$ 24 C {{d a(t) }\over {dt}} +12 C^2 \Lambda a^3 +1 =0 . \eqno {29} $$

We could not solve this equation for $a(t)$ as a function of $t$ analytically; 
actually we could not invert the solution we got for $t$ as a function of $a$. 
We give the numerical solution below. (See Figure) We see that asymptotically  $ a(t)
^3 = {{M^3}\over {4}} $. In this region the scalar field is given as 
$$ \phi= \sqrt{ {{1}\over { \pi( (16)^{1/3} \sqrt {3M\Lambda} r+1) }}} . \eqno 
{30} $$

We can also couple a complex field to the gravitational field in the presence of a constant  potential.  Nonzero solutions both for the real and the imaginary parts of the scalar field can be found.

If we try a rotating black-hole interacting with a scalar field, we start with
$$ ds^2=-v(r) dt^2+w(r)dr^2+r^2d\theta^2+2Jdtd\theta . \eqno {31} $$
Here 
$$ v(r)= \Lambda r^2-M , \eqno {32} $$
$$ w(r) = {{r^2(2\Lambda r^2 -M)^{{{3}\over{2}}}}\over{(J^2+\Lambda r^4 -Mr^2)(2c_{1} 4((2\Lambda r^2 -M)^{{{3}\over{2}}})}}, \eqno {33} $$
$$ \phi(r) = \sqrt{ {{-8+\sqrt{{{ (2\Lambda r^2 -M)^{{{3}\over{2}}}}\over{c_{2}(2c{1}+(2\Lambda r^2 -M)^{{{3}\over{2}}}}}}}}} \eqno{34} $$
Here 
$$c_{1}= {{1}\over{2}} (M^2-4J^2\Lambda)^{{3}\over{4}}, \eqno {35} $$
$$c_{2} = {{1}\over{64}} . \eqno {36} $$
We choose the constants $ c_{1}, c_{2} $ so that the scalar field goes to zero as $r$ goes to infinity and there are two singular points given by the inner and outer horizons
$$ r_{-}= \sqrt{{{M-\sqrt{M^2-4J^2\Lambda}}\over {2\Lambda}}},  $$
$$ r_{+}=\sqrt{{{M+\sqrt{M^2-4J^2\Lambda}}\over {2\Lambda}}}. \eqno {37} $$

Here between the inner and outer horizons, the scalar field is undefined. We use the solution when $r> r_{+} $. 
\vskip.5truein
\noindent
{\bf{
THERMODYNAMICS}}
\vskip.2truein
We will use the so called KKW method [10,11,12] to calculate the entropy of the 
solutions above. 
We first employ a transformation, the so-called Painlev\' e transformation, to 
get rid of the coordinate singularities in the metric. 

For a metric of the form 
$$ ds^2= -F(r) dt^2 + F(r)^{-1} dr^2 + r^2 d\theta^2 \eqno {38} $$
we set 
$$ \sqrt { F(r)} dt = \sqrt {F(r)} d\tau - {{\sqrt{1-F(r)}}\over {F(r)}} dr , 
\eqno {39} $$
which reduces the metric to the form
$$ ds^2 = -F(r) d\tau^2 +dr^2 + 2 \sqrt{ 1-F(r)} dr d\tau +r^2 d\theta^2 . 
\eqno {40} $$
To find the black body radiation, we take 
$ds=0$, $ d\theta=0$, which gives us the equation of motion for the outgoing 
particles.
$$ {{dr}\over{dt}}= 1-\sqrt{1-F(r)} \eqno {41} $$
Since this particle takes its energy $\omega$ from the black-hole, the mass of 
the black-hole is reduced to $M-\omega$.
We write equation 41 as 
$$ {{dr}\over{dt}}= 1-\sqrt{1-F(r,M-\omega)} . \eqno {42} $$
We define the change in entropy of the black-hole as
$$\Delta S_{BH} = S_{BH}(M-\omega) - S_{BH}(M) ,\eqno{43} $$ where the 
imaginary part of the action $\it{S}$ is related to the change in entropy as
$$ -2Im{\it{S}} =\Delta S_{BH}. \eqno {44} $$ 
The imaginary part of the action is calculated as
$$ Im{\it{S}}= \int_{r_{+}(M)}^{r_{+} (M-\omega)} \int_{0}^{\omega} {{d\omega' 
dr}\over{{{dr}\over{dt}}(M,M-\omega')}}. \eqno {45} $$
The tunnelling probability is expressed as
$$ \Gamma= \exp(\Delta S_{BH}) . \eqno {46}$$
Using our solutions we set
$$ Im{\it{S}}= \int_{r_{+}(M)}^{r_{+} (M-\omega)} \int_{0}^{\omega} {{d\omega' 
dr}\over{1-\sqrt{1-\Lambda r^2+(M-\omega')}}}. \eqno {47} $$
where $r_{+}= \sqrt{{{M}\over{\Lambda}}}$. Contour integration gives 
$$ Im{\it{S}} =2\pi \left(\sqrt{{{M}\over{\Lambda}}} - \sqrt{{{M-\omega}\over
{\Lambda}}} \right)\eqno {48} $$
which gives the black-hole entropy
$$S_{BH} = 4\pi \sqrt{{{M}\over{\Lambda}}} \eqno {49} $$ 
and tunnelling probability
$$ \Gamma = \exp { 4\pi \left(-\sqrt{{{M}\over{\Lambda}}} + \sqrt{{{M-\omega}
\over{\Lambda}}} \right)}. \eqno {50} $$
Using the definition 
$$\Gamma= \exp(\Delta S_{BH})= \exp( - {{\omega}\over{T}}) \eqno {51}$$
for the Hawking temperature of the black-hole, we find 
$$ T= {{\omega}\over{4\pi}} \left(\sqrt{{{M}\over{\Lambda}}} - \sqrt{{{M-\omega}
\over{\Lambda}}} \right)^{-1}, \eqno {52} $$
which gives the Hawking temperature 
$$T_{H} = {{\sqrt{M \Lambda}}\over {2\pi}} \eqno {53} $$
upon expanding around $\omega=0$.

Same expressions can be found [8] using the relation between black-holes and 
thermodynamics which will also give 
$$ S_{BH} = 4\pi r_{+} . \eqno {54} $$
For the rotating black-hole case this expression gives 
$$ S_{BH}= 4\pi \sqrt{{{ M+\sqrt{M^2 - 4\Lambda J^2}}\over{2\Lambda}}} \eqno 
{55} $$
with 
$$T= {{\omega}\over{ S_{BH}(M)-S_{BH}(M-\omega)}} . \eqno {56} $$
Expanding around $\omega$ equals zero gives 
$$ T_{H} = {{\sqrt{M^2-4J^2 \Lambda}}\over{\sqrt{2} \pi \sqrt{{{M+\sqrt{M^2-4
\Lambda J^2}}\over{\Lambda}}}}}. \eqno {57} $$

If we couple the gravity with a scalar field, our expression for $F(r)$ in eq.
(40) is written as
$$ F(r)= \Lambda r^2 - {{2 \sqrt{M^3}}\over{3\sqrt{3\Lambda} r}} -M . \eqno 
{58} $$
which gives 
$$ S_{BH} = {{16\pi}\over{3\sqrt{3\Lambda}}} \sqrt {M} \eqno {59} $$
$$ T(\omega) = {{3\omega\sqrt{3\Lambda}}\over{16\pi}} (\sqrt{M}-\sqrt{M-\omega})
^{-1}. \eqno {60} $$
Expansion around $\omega$ equals zero gives 
$$ T_{H}= {{3\sqrt{3M\Lambda}}\over{8\pi}} , \eqno {61} $$
which is the same expression as the one derived from purely thermodynamical 
expressions [8].

Note that in equation 27, $-a(t)$ has taken the place of $M$ in the solutions without the scalar field, equation 9.  If the action $A$ is calculated, we find that $A={{M}\over{T}}$ where $T$ equals the Hawking temperature. Thus,   $M$  is related to the energy of the black-hole, and as time elapses,  $-a(t)$ decreases,(See Figure) ,  showing that the energy  of the black-hole is decreasing as well.

\vskip.5truein
\vskip.5truein
\noindent
{\bf{
NEWMAN-PENROSE COEFFIENTS}}
\vskip.2truein
We can calculate the coefficients for these solutions and show how they differ 
from those for the latter solution given in reference [7]. The metric of 
reference [7] is given as 
$$ ds^2=-(N^2F^2-R^2N^{\phi 2})dt^2+2R^2 N^{\phi} d\theta dt +R^2d\theta^2 + F^
{-2} dR^2 . \eqno {62} $$
Here
$$f^2= r^2-\tilde{M}-{{1}\over{4}}\tilde{Q}^2\log r^2, \eqno {63} $$
$$ R^2 = {{ r^2-\omega^2 f^2}\over{1-\omega^2}}, \eqno {64} $$
$$ F^2 = \left( {{dR}\over{dr}}\right)^2 f^2, \eqno {65} $$
$$ N= {{r}\over {R}} \left( {{dr}\over {dR}}\right), \eqno {66} $$
$$ N^{\phi} = {{\omega (f^2-r^2)}\over {(1-\omega^2)R^2}} .\eqno {67} $$
where $\tilde{M} = {{(1-\omega^2)M}\over{1+\omega^2}},  \tilde{Q}=\sqrt{1-
\omega^2} Q $. $\omega$ is related to the angular velocity related to rotation. 
The cosmological constant is negative and set to unity.

One may use the Newman-Penrose formalism for 2+1 dimensions as given Aliev and 
Nutku [14]. Similar formalism is also given in references [15].

In this formalism the metric is written in terms of basis 1-forms
$$ ds^2 = l \times n + n \times l - m \times m . \eqno {68} $$
The Ricci rotation coefficients are defined as 
$$ dl = -\epsilon l \wedge n +(\alpha - \tau) l \wedge m - \kappa n \wedge m , 
\eqno {69} $$
$$ dn = \epsilon' l \wedge n - \kappa' l \wedge m -(\alpha+ \tau') n \wedge m , 
\eqno {70} $$
$$ dm = (\tau' - \tau) l \wedge n - \sigma' l \wedge m - \sigma n \wedge m . 
\eqno {71} $$
Connection 1-forms, which give the spin coefficients, are defined as
$$ \Gamma_{0}^{0} = {{1}\over {2}} ( -\epsilon' l + \epsilon n + \alpha m ) , 
\eqno {72} $$
$$ \Gamma_{0}^{1} = -{{1}\over {\sqrt{2}}} (-\tau l - \kappa n + \sigma m) , 
\eqno {73} $$
$$ \Gamma_{1}^{0} = {{1}\over {\sqrt{2}}}( -\kappa' l - \tau' n + \sigma' m) . 
\eqno {74} $$
The 2-forms are obtained from the equation 
$$ R_{a}^{b} = d\Gamma_{a}^{b}- \Gamma_{a}^{m} \wedge \Gamma _{m}^{b} . \eqno 
{75} $$ 
The curvature 2-form is written in terms of the basis 2-forms, written in terms 
of the triad scalars:
$$ R_{0}^{0} = \left( 2 \Phi_{11} -{{3}\over{2}} \Lambda \right) l \wedge n - 
\Phi_{12} l \wedge m + \Phi_{10} n \wedge m , \eqno{76} $$
$$ R_{0}^{1} = \sqrt{2} \Phi_{01} l \wedge n + {{1}\over{\sqrt{2}}} \Phi_{02} l 
\wedge m - \sqrt {2} \Phi_{00} n \wedge m , \eqno {77} $$
$$ R_{1}^{0} = \sqrt{2} \Phi_{12} l \wedge n - \sqrt{2} \Phi_{22} l \wedge m + 
{{1}\over{\sqrt{2}}} \Phi_{02} n \wedge m . \eqno {78} $$
Here curvature scalar $R$ is defined as $ \Lambda = {{1}\over {18}} R $. 

Rewriting the metric in terms of 
$$ q(r) = \left[ r^2 f^2- {{\omega^2}\over {(1-\omega^2)^2 }} (f^2 - r^2)^2
\right], \eqno{79 } $$
$$ k(r) = -{{2}\over {\omega}} (R^2 - r^2) , \eqno {80 } $$
we obtain
$$ ds^2 = -{{q(r)}\over {R^2 }}dt^2+k(r) d\theta dt + R^2 d\theta^2 + F^{-2} 
dR^2. \eqno { 81 } $$
We define the triad system as
$$ l = {{1}\over{\sqrt{2}}} [(k-2rf)dt+2R^2 d\theta], \eqno {82} $$
$$ n = {{1}\over{\sqrt{2}}}\left[{{(k+2rf)}\over{4R^2}} dt+ {{1}\over{2}} 
d\theta \right], \eqno { 83 } $$
$$ m = {{1}\over{\sqrt{2}}} F^{-1} dR . \eqno {84 } $$

One can calculate the basis 2-forms straightforwardly and obtain:
$$\Phi_{01}=\Phi_{10}=\Phi_{12}=0 , \eqno {85}$$
$$ \Phi_{02} = {{1}\over{\sqrt{2}}}, \eqno {86} $$
$$ \left( 2 \Phi_{11} -{{3}\over{2}} \Lambda \right)= \left({{\tilde{Q^2}}\over
{8r^2}} -{{1}\over{2}}\right), \eqno {87} $$
$$ \sqrt{2}\Phi_{00} = \tilde{Q}^2\left({{\omega^2 }\over{(1-\omega^2)}}+1 -
{{\omega f}\over{(1-\omega^2)r}}-{{R^2}\over{2r^2}}\right) , \eqno {88} $$ 
$$\sqrt{2} \Phi_{22}= \left( {{1}\over{2R^2}}+ {{\tilde{Q}^2 (r+\omega f)}\over
{16(1-\omega^2)rR^4 }}-{{3\tilde{Q}^2}\over{32 r^2 R^2 }} \right) . \eqno {89} 
$$

If we use the "wrong solution", as given in eq. (16),
$$ ds^2 =-\left( v( r ) + {{J ( r )^2}\over {r^2}} \right) dt^2 + {{1}\over{v( 
r )}}dr^2 + \left( rd\theta +{{J( r ) }\over{r}}dt \right)^2 , \eqno {16} $$
where the parameters used  are defined in equations (17) and (18), and 
calculate the coefficients given above, we find that the same Ricci rotation 
coefficients and the same triad scalars, $\Phi_{ij} $ vanish. The non-vanishing 
terms are different algebraically though, equality being established only when 
$J,Q$ and $\omega$ vanish.
We do not give the details of this straightforward calculation. We just give 
the result of the scalar triad corresponding to the one given in equation  79 
above.

For the metric in eq. (16), we get 
$$\left( 2 \Phi_{11} -{{3}\over{2}} \Lambda \right)= -\left(2-{{2B^2}\over{r^2}}-
{{2J^2}\over{r^4}} \right)\left({{1}\over{4}}-{{{{1}\over{4}}{{J^2}\over{r^2}}}
\over{4 r^2 +{{5J^2}\over{r^2}} -4M-8 B^2 log{{r}\over{r_{+}}}}}\right). \eqno
{90}$$
\vskip.5truein
\noindent
{\bf{CONCLUSION}}
\vskip.2truein
In this work  we derived black hole solutions in three dimensionswith and without interacting  with a scalar field. Except for
 the case where the scalar particle is time dependent, these solutions are 
found in the literature,[2,7,8].  We then calculated the  entropies of the different black holes, 
the Hawking temperatures and the emission probabilities. We found that even for the case where the scalar field is time dependent, the mass of the black-hole goes to a constant. The value of the constant is less than the value at time equals  zero, which shows that energy is carried away.   Finally we calculated 
the Newman-Penrose coefficients for the charged and rotating solution and 
compared these coefficients  with those of the "wrong" solution. We found that in both cases the 
null ones are the same, but  the algebraic expressions for the non-zero 
coefficients differ. 
\vskip.5truein
\noindent
{\bf{ Acknowledgement:}} This work is partially supported by 
 by T\" UBITAK , the Scientific and Technical Research 
Council of Turkey. M.H. is also supported by TUBA, the Academy of Sciences of 
Turkey.
\vfill\eject
\noindent
{\bf{REFERENCES}}
\vskip.2truein

\item {1.} Abdalla,E, M.C.B. Abdalla, K.D.Rothe, 1991, {\it{2 Dimensional 
Quantum Field Theory}}, World Scientific, Singapur ; Zinn-Justin, J, 1993,{\it
{Quantum Field Theory and Critical Phenomena}}, Clarendon Press, Oxford,UK.

\item {2.} M.Banados, C. Teitelboim and J.Zanelli, Phys. Rev. Lett. {\bf{69}} 
(1992) 1849; also M.Banados, M.Henneaux, C. Teitelboim and J.Zanelli, Phys. 
Rev. D {\bf{48}}(1993) 1506;

\item {3.} S.Deser, R. Jackiw and S.Templeton, Ann. Phys.(NY){\bf{140}} (1982) 
372; S. Deser, R. Jackiw and G.'t Hooft, Ann. Phys. (NY){\bf{152}} (1984) 220; 
S.Deser and R. Jackiw, Ann. Phys. (NY) {\bf{153}} (1984) 405;

\item {4.} G.Cl\' ement, Class. Quantum Grav. {\bf{10}} (1993) L54;

\item {5.} M.Kamara and T.Koikawa, Phys. Lett. B {\bf{533}} (1995) 196; 
M.Cataldo and P.Salgado, Phys. Lett B {\bf{448}} (1999) 20;

\item {6.} A.Garcia, hep-th/990911, S.Fernando and F.Mansoori, gr-qc/9705016;

\item {7.} C.Martinez, C. Teitelboim and J.Zanelli, Phys. Rev. D {\bf{61}} 
(2000) 104013;

\item {8.} C.Martinez and J.Zanelli, Phys. Rev.D {\bf{54}} (1996) 3830; M.Hennaux, C.Martinez, Ricardo Troncoso and J.Zanelli, Phys. Rev. D 
{\bf{65}} (2001) 104007;

\item {9.} K.S. Virbhadra,
    Pramana, {\bf {44}} (1995) 317; Oscar J.C. Dias and Jose P.S. Lemos, JHEP 01 (2002) 006, Phys.Rev. D {\bf{64}} (2001) 064001; Phys. Rev D {\bf{66}} (2002) 024034;

\item {10.} P.Kraus and F.Wilczek, Nucl. Phys. B {\bf{433}} (1995) 403;

\item {11.} P.Kraus, F. Wilczek, Nucl. Phys.B {\bf{437}} (1995) 231;

\item {12.} E.Keski-Vakkuri and P.Kraus, Nucl. Phys. B {\bf{491}} (1997) 249;

\item {13.} E.T.Newman and R.Penrose, J. Math. Phys. {\bf{3}} (1962) 566, {\bf
{4}} (1962) 998;

\item {14.} A.N.Aliev and Y.Nutku, Class. Quantum Grav. {\bf{12}} (1995) 2913, 
also gr-qc/9812090;

\item {15.} G.S.Hall, T. Morgan and Z.Perjés, Gen. Rel. Grav.,{\bf{19}} (1987) 
1137; O.Dreyer, Penn. State Ph.D.Dissertation (2001), Appendix A
\vfill\eject
{\bf{ FIGURE CAPTION}}\vskip .2truein
\noindent
The plot of $a(t)$ vs. time for the time dependent scalar field when $M=10, \Lambda=0.01$
\vfill\eject
\midinsert
\vskip .75truein
\centerline{\BoxedEPSF{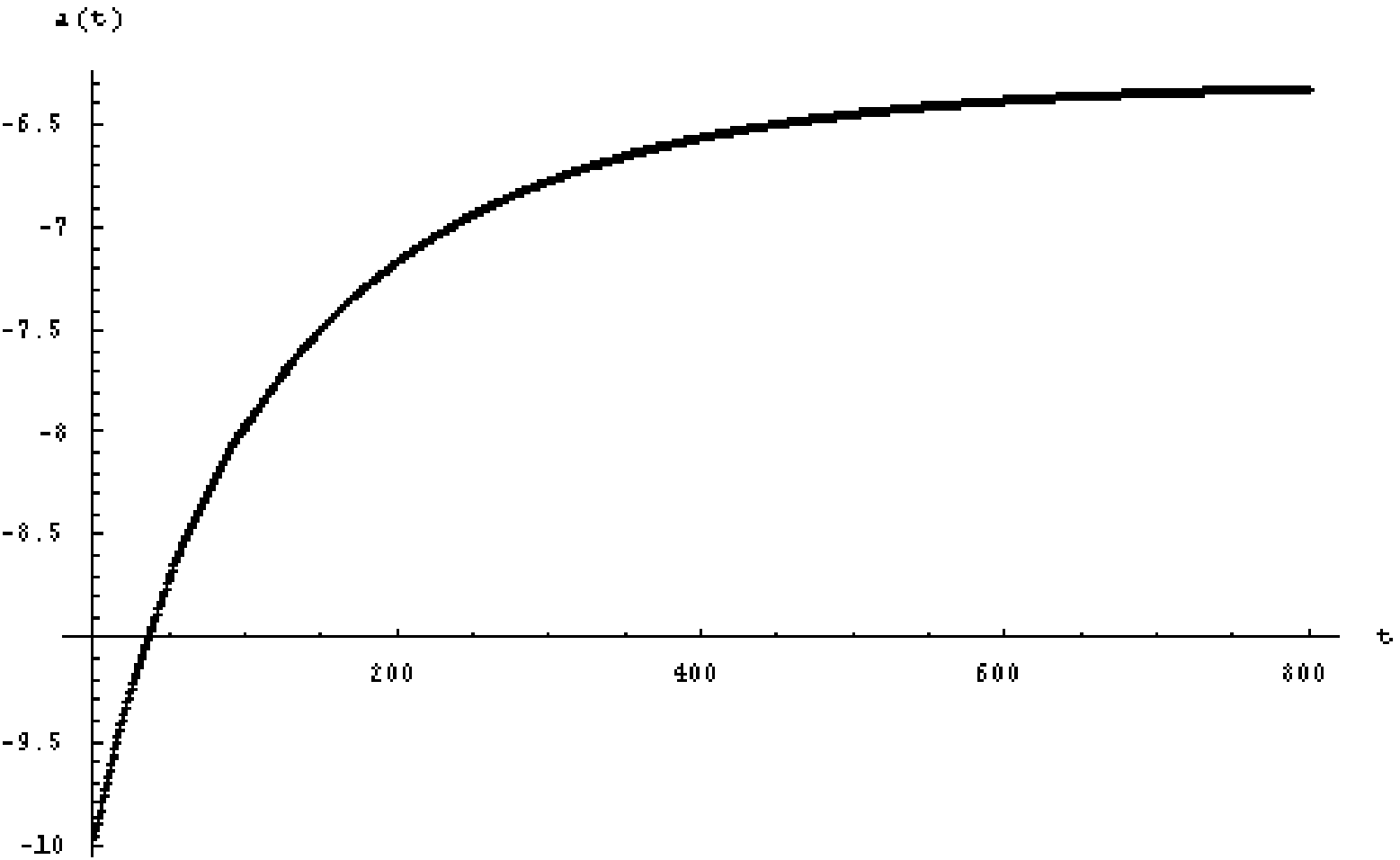}}
\centerline {\bf { Figure }}
\endinsert
\end